\documentclass[11pt]{article}
\PassOptionsToPackage{dvipsnames,svgnames}{xcolor}
\usepackage[preprint]{acl}

\usepackage{times}
\usepackage{adjustbox}
\usepackage{latexsym}
\usepackage{booktabs}
\usepackage{multirow}
\usepackage{tabularx}
\usepackage{placeins}
\usepackage{float}
\usepackage[T1]{fontenc}
\usepackage{amsfonts}
\usepackage{amssymb}
\usepackage{amsmath}

\usepackage[utf8]{inputenc}

\usepackage{microtype}

\usepackage{inconsolata}
\usepackage{xcolor}         
\usepackage{graphicx} 
\usepackage[most,breakable]{tcolorbox}

\usepackage[normalem]{ulem}
\usepackage{enumitem}
\usepackage{color}
\usepackage{hyperref}
\useunder{\uline}{\ul}{}
\definecolor{gold}{RGB}{255,215,0}
\definecolor{silver}{RGB}{192,192,192}
\definecolor{bronze}{RGB}{205,127,50}

\definecolor{boxcolor1}{RGB}{235,240,255} 
\definecolor{boxcolor2}{RGB}{255,240,230} 
\definecolor{boxcolor3}{RGB}{230,255,230} 
\definecolor{boxcolor4}{RGB}{255,230,255} 
\definecolor{boxcolor5}{RGB}{255,255,230} 
\definecolor{boxcolor6}{RGB}{230,255,255} 
\definecolor{grey}{RGB}{240,240,240} 

\newtcolorbox{promptbox}{
  colback=gray!5,
  colframe=black,
  boxrule=0.6pt,
  arc=0pt,
  left=5pt,
  right=5pt,
  top=5pt,
  bottom=5pt,
  enhanced
}

\title{AstroSpecLM: A Spectrum-Language Model for Evidence-Grounded Astronomical Spectral Analysis}

\author{Jinghang Shi, Yanxia Zhang, Ali Luo, Changhua Li and Xiao Kong }

\begin{document}
\maketitle

\begin{abstract}
Astronomical spectra encode rich physical information, but drawing scientific conclusions from spectral features typically requires expert interpretation. This paper presents AstroSpecLM, a spectrum-language model that connects one-dimensional DESI spectra with Qwen3-4B to answer questions and provide explanations grounded in spectral evidence. Instead of generating question-answer pairs directly from templates or raw catalog fields, we first distill each spectrum into a compact set of catalog- and spectrum-derived facts, then use these facts as references to generate instruction-following conversations. The resulting model is competitive with specialist supervised baselines on classification and redshift estimation, while additionally producing natural-language explanations that reference specific spectral features. Our results indicate that grounding a language model in one-dimensional scientific spectra is feasible, and that fact-mediated instruction data yields a model capable of both prediction and explanation.
\end{abstract}

\section{Introduction}
\label{sec:introduction}
Astronomical spectra provide wavelength-resolved measurements that support the inference of object type, redshift, stellar temperature, and other physical properties. A typical astronomical spectrum consists of a one-dimensional (1D) flux array spanning thousands of wavelength pixels. Its interpretation hinges on linking localized spectral features, such as emission and absorption lines, and their wavelength shifts to the physical properties of the astronomical source. This process is inherently evidence-based: conclusions about an object's type or redshift should be supported by specific spectral features. Modern spectroscopic surveys such as DESI produce millions of such spectra \citep{desicollaboration2024edr,desi2026dr1}, and automated pipelines can efficiently extract structured measurements from them \citep{guy2023desipipeline}. However, the development of natural-language frameworks that explicitly map this spectral evidence to final conclusions lags behind. Ideally, such a model should process spectral inputs to generate interpretable predictions, accompanied by a clear rationale grounded in the underlying spectral features.

Large language models (LLMs) offer a path toward this form of grounded scientific interaction \citep{taylor2022galactica,zhang2024scientificllmsurvey}. When connected to modality encoders, they can condition generation on non-text inputs and produce flexible natural-language responses \citep{alayrac2022flamingo,li2023blip2}. However, most multimodal language models are designed for natural images \citep{liu2023llava,zhu2024minigpt4}, videos \citep{maaz2024videochatgpt,lin2024videollava}, or documents \citep{hu2025docowl2}, where the input is two-dimensional and semantically compositional. A 1D astronomical spectrum poses a different grounding problem: its meaning is tied to wavelength, and critical evidence often lies in narrow localized regions that only become meaningful through physical interpretation \citep{guy2023desipipeline,desicollaboration2024edr,zhong2024gasnet,parker2024astroclip}. Bridging this gap requires both a spectrum encoder that preserves wavelength-localized information and a training strategy that grounds language-model outputs in observable spectral evidence \citep{he2022mae,parker2024astroclip}.

In this work, we present \textbf{AstroSpecLM}, a spectrum-language model that connects real 1D DESI spectra with Qwen3-4B \citep{yang2025qwen3} for spectrum-grounded question answering. AstroSpecLM uses a spectrum-specific masked autoencoder \citep{he2022mae}, \textbf{SpecMAE}, to encode the input spectrum, and a lightweight spectral projector to insert spectral representations into the language-model embedding space. To train the model faithfully, we construct \textbf{DESI-SpecInstruct}, an instruction-tuning dataset built through an intermediate fact extraction stage. The resulting model answers spectrum-grounded questions in natural language with explanations that reference specific spectral features as evidence.

\section{Related Work}
\label{sec:related-work}
Astronomical spectral analysis is an evidence-centered form of scientific interpretation: physical conclusions are supported by continuum shapes, emission and absorption features, spectral breaks, and wavelength shifts. Modern spectroscopic pipelines and expert systems have made this process increasingly scalable, producing calibrated spectra together with structured measurements and quality assessments \citep{guy2023desipipeline,desi2026dr1}. Recent spectral models further move from task-specific prediction toward reusable representations, including image--spectrum contrastive models, stellar spectral foundation models, and native-resolution spectral encoders \citep{parker2024astroclip, koblischke2024spectrafm,parker2025aion, islam2026omnispectra}. These systems are effective for processing spectra at scale, but their outputs --- structured labels, numerical estimates, or visual interfaces --- leave the evidence-to-conclusion chain implicit and dependent on expert interpretation.

LLMs have transformed the interface of AI systems from fixed-output prediction to open-ended instruction following and natural-language interaction \citep{ouyang2022instructgpt, chung2024flan, grattafiori2024llama3, yang2025qwen3}. Multimodal LLMs extend this capability to non-text inputs by connecting modality encoders with language models through adapters, projection layers, or cross-attention mechanisms \citep{liu2023llava, li2024llavaonevision, dai2024nvlm, zhu2025internvl3, bai2025qwen3vl}. These advances suggest a general path for turning perceptual signals into language-conditioned systems. However, most general-purpose multimodal language models are optimized for natural images, videos, documents, or web-scale data, rather than real 1D scientific measurements whose interpretation depends on wavelength-localized physical evidence such as emission line profiles and continuum breaks.

Within astronomy, this language-centered paradigm has been adopted through domain-specific LLMs, astronomical vision-language models, and early spectroscopic LLM systems. Astronomy-oriented LLMs improve access to domain knowledge by continued pretraining or instruction tuning on astronomical literature and question-answer data \citep{nguyen2023astrollama,dehaan2025astrosage}. Astronomical multimodal models further connect observations with language, including image--text retrieval, astronomical visual question answering, and radio-source analysis \citep{mishrasharma2024paperclip,zaman2025astrollava,riggi2025radiollava}. More closely related to our setting, recent work has begun to bring language-model interfaces into spectroscopic workflows: Teaching LLMs to Speak Spectroscopy adapts LLaMA-3.1-8B with LoRA to predict galaxy redshifts from spectroscopic data while retaining part of its language ability \citep{ramachandra2025speak}, and Spec-o3 develops a tool-augmented vision-language agent for rare-object vetting through spectral-image inspection and interleaved reasoning \citep{jia2026speco3}. These studies highlight the promise of bringing language-model interfaces to spectroscopic workflows, but they are still closer to limited quantitative prediction or tool-mediated visual inspection. AstroSpecLM bridges spectral expert systems and astronomy-oriented language models. It uses real observed 1D spectra as the direct evidence source, supporting unified classification, regression, spectral-feature analysis, and evidence-based explanation within a single generative framework.

\section{Data Construction}
\label{sec:data}
\begin{figure*}[t]
    \centering
    \includegraphics[width=0.93\linewidth]{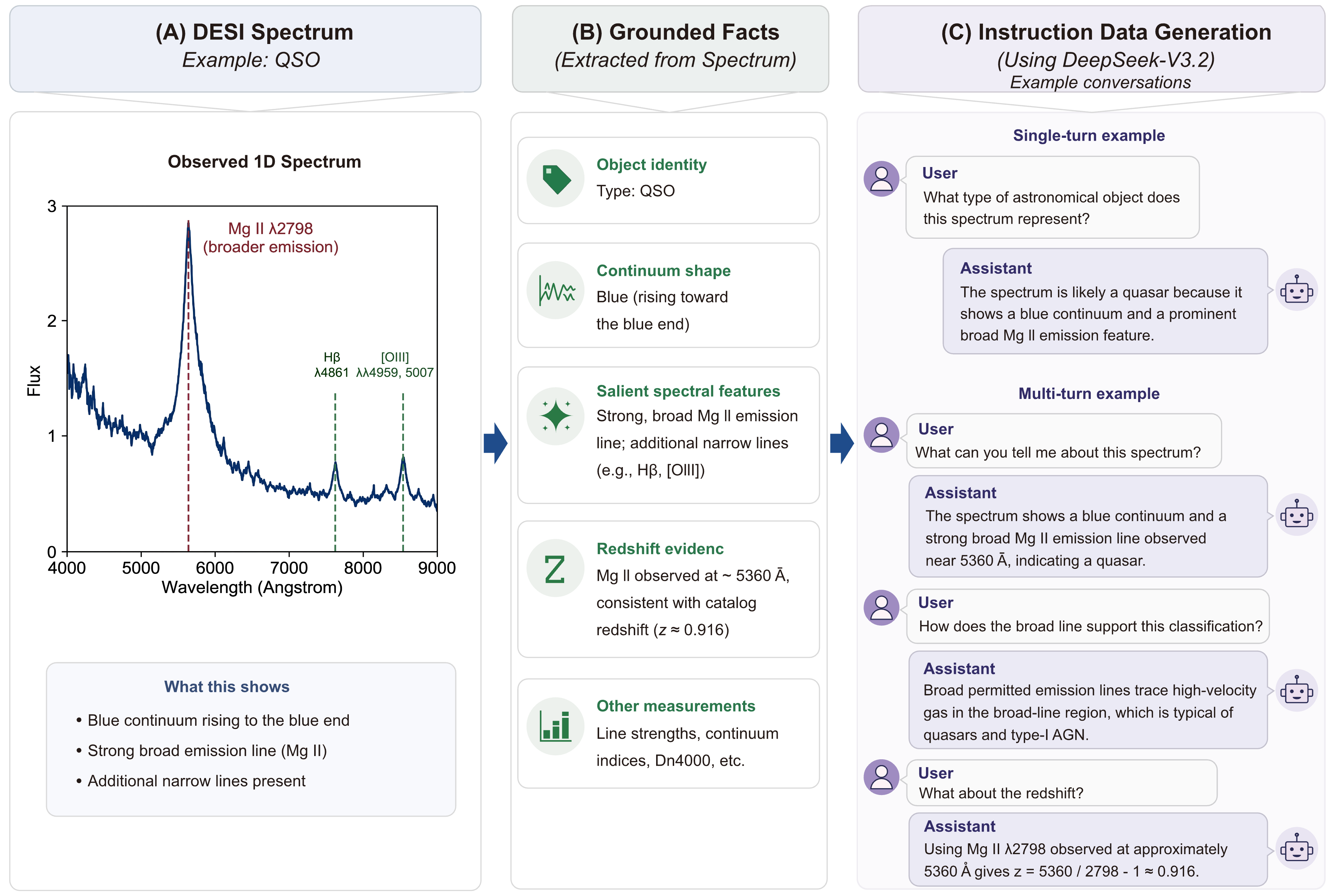}
    \caption{Instruction data construction pipeline. Raw DESI spectra are first distilled into grounded facts combining catalog annotations and spectral measurements; these facts then serve as references for generating instruction-following conversations.}
    \label{fig:data-pipeline}
\end{figure*}

A core challenge in spectrum-grounded question answering is encouraging generated answers to remain supported by observable spectral evidence. To this end, we design a pipeline that converts DESI spectra into language supervision through an intermediate \emph{fact extraction} stage, producing the DESI-SpecInstruct dataset. Figure~\ref{fig:data-pipeline} illustrates the process: we first select high-quality spectra, distill each into a compact set of catalog- and spectrum-derived facts, and then use these facts as grounding references to generate instruction data. This design reduces the risk that generated answers introduce unsupported physical properties, while enabling rich, evidence-tied questions and explanations.

\subsection{Spectrum Selection}
\label{sec:spec-selection}
We use 1D spectra from DESI DR1 \citep{desi2026dr1,guy2023desipipeline}, defining two data sources for different training stages. The pretraining corpus for the spectral encoder, denoted as \(\mathcal{D}_{\mathrm{enc}}\), comprises 2.85 million spectra balanced across STAR, GALAXY, and QSO classes, and is used for masked reconstruction training. From the same underlying DESI collection, we further select a higher-quality subset \(\mathcal{S}_{\mathrm{inst}}\) of 58,558 spectra with reliable catalog annotations and clean spectral measurements for instruction-data construction. Detailed acquisition, preprocessing, and filtering criteria are provided in Appendix~\ref{app:corpus-details}.

\subsection{Fact Extraction}
\label{sec:fact-extraction}
For each spectrum in \(\mathcal{S}_{\mathrm{inst}}\), we construct a set of \emph{grounded facts} that combine catalog labels (e.g., object class, redshift) with quantities directly measured from the spectrum (e.g., emission-line equivalent widths, continuum slopes). These facts are never shown to the model; they act solely as grounding references during instruction data generation, ensuring that every generated question and answer is anchored to spectral evidence. A full specification of fact types and measurement rules is given in Appendix~\ref{app:fact-details}.

\subsection{Instruction Data}
\label{sec:instruction-data}
Using the grounded facts as a conditioning signal, we employ DeepSeek-V3.2 \citep{deepseek2025v32} to generate user--assistant conversations, following the broader practice of constructing instruction data with LLMs \citep{wang2023selfinstruct,ding2023ultrachat}. Both single-turn and multi-turn interactions are produced. To improve faithfulness to the input spectrum, the generation prompt explicitly forbids the model from introducing wavelengths, redshifts, or physical quantities not supported by the provided facts, thereby reducing the risk of hallucinated measurements. The full prompt is listed in Appendix~\ref{app:prompt-details}.

After automatic filtering for malformed conversations and leakage of reference wording, DESI-SpecInstruct contains 305,536 conversations (208,614 single-turn and 96,922 multi-turn).

\subsection{Held-out Benchmark}
\label{sec:benchmark}
We construct AstroSpecQA as a held-out benchmark from spectra disjoint from those used in DESI-SpecInstruct. It contains 11,698 questions across four spectrum-grounded tasks: object classification (5,000), redshift regression (3,247; galaxies and QSOs only), stellar subtype classification (1,753; stars only), and effective-temperature regression (1,698; stars only). This benchmark is used only for evaluation, and its label distributions are reported in Appendix~\ref{app:benchmark}.

\section{Method}
\label{sec:method}
\begin{figure*}[t]
    \centering
    \includegraphics[width=0.91\linewidth]{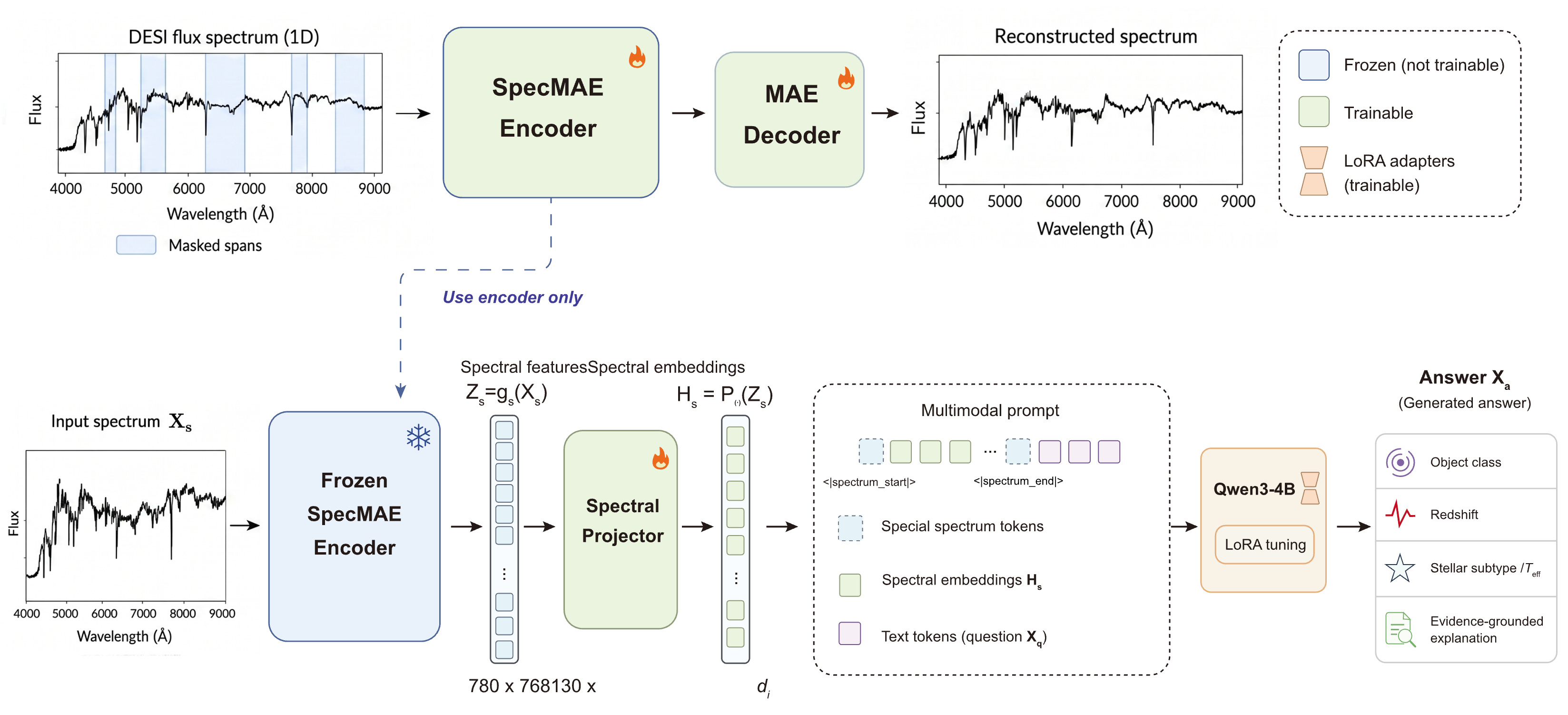}
    \caption{Overview of AstroSpecLM. A normalized DESI spectrum is encoded by the SpecMAE encoder, compressed by a spectral projector, and inserted into Qwen3-4B as continuous spectral embeddings. During spectral instruction tuning, the SpecMAE encoder is frozen, while the projector, special-token embeddings, and LoRA adapters are trained.}
    \label{fig:astrospeclm}
\end{figure*}

AstroSpecLM is a spectrum-language model built on Qwen3-4B and a spectrum-specific masked autoencoder, SpecMAE. It connects 1D astronomical spectra to the language model by projecting spectral representations into the Qwen3-4B embedding space, and is trained through a two-stage procedure: spectrum masked autoencoding followed by spectral instruction tuning. Figure~\ref{fig:astrospeclm} illustrates the overall architecture.

\subsection{AstroSpecLM}
\label{sec:astrospeclm}
 AstroSpecLM takes a normalized spectrum \(\mathbf{X}_s \in \mathbb{R}^{7781}\) and a natural-language question \(\mathbf{q}\) as input, and generates an answer \(\mathbf{a}\) conditioned on both the spectral signal and the textual instruction. The spectrum is first encoded by a pretrained spectral encoder and then projected into the language-model embedding space.

Given a spectrum \(\mathbf{X}_s\), the SpecMAE encoder \(g_s(\cdot)\) produces a sequence of spectral patch representations:
\begin{equation}
    \mathbf{Z}_s = g_s(\mathbf{X}_s) \in \mathbb{R}^{778 \times 768}.
\end{equation}
These features are passed to a lightweight spectral projector \(P_{\phi}(\cdot)\). To reduce the spectral sequence length before insertion into the language model, we pad \(\mathbf{Z}_s\) from 778 to 780 tokens and merge every six adjacent spectral tokens, producing 130 spectral tokens. The merged sequence is mapped to the hidden dimension of Qwen3-4B (\(d_{\ell}=2560\)):
\begin{equation}
    \mathbf{H}_s = P_{\phi}(\mathbf{Z}_s) \in \mathbb{R}^{130 \times d_{\ell}}.
\end{equation}

The projected spectral embeddings are inserted into the language-model input between two newly added special tokens, \texttt{<|spectrum\_start|>} and \texttt{<|spectrum\_end|>}. Let \(E(\mathbf{q})\) denote the token embeddings of the question, and let \(\mathbf{e}_{\mathrm{start}}\) and \(\mathbf{e}_{\mathrm{end}}\) denote the embeddings of the two special tokens. The input embedding sequence is
\begin{equation}
    \mathbf{E}_{\mathrm{in}} =
    [\mathbf{e}_{\mathrm{start}}, \mathbf{H}_s,
    \mathbf{e}_{\mathrm{end}}, E(\mathbf{q})].
\end{equation}

The language model then generates the answer autoregressively:
\begin{equation}
    p(\mathbf{a}\mid \mathbf{X}_s,\mathbf{q})
    =
    \prod_{i=1}^{L}
    p_{\theta}(a_i \mid \mathbf{E}_{\mathrm{in}}, a_{<i}),
\end{equation}
where \(\mathbf{a}=(a_1,\ldots,a_L)\) is the target answer and \(\theta\) denotes the trainable parameters during spectral instruction tuning.

\subsection{Spectrum Masked Autoencoding}
\label{sec:specmae}

We pretrain SpecMAE on \(\mathcal{D}_{\mathrm{enc}}\) (Section~\ref{sec:spec-selection}) using a masked reconstruction objective. Many diagnostic spectral features are spectrally localized; we therefore mask continuous spectral spans rather than isolated tokens, encouraging the encoder to infer missing regions from surrounding wavelength context. Specifically, for each spectrum, we sample 10 masked chunks, each containing 30 consecutive patch tokens.

The encoder first applies a 1D convolutional patch embedding with kernel size 20 and stride 10, producing overlapping spectral patches with boundary padding. This results in 778 patch tokens, which are processed by an 8-layer Transformer encoder \citep{vaswani2017attention} with hidden size 768, 8 attention heads, MLP ratio 4, and learned positional embeddings.

During pretraining, the encoder observes only the visible patch tokens, while a lightweight decoder reconstructs the masked flux values. The decoder is discarded after pretraining. The reconstruction loss is computed only on wavelength pixels covered by masked patches. Let \(f_i\) be the original flux at pixel \(i\), \(\hat{f}_i\) the reconstructed flux, \(m_i\in\{0,1\}\) the strict mask indicator, and \(v_i\in\{0,1\}\) a validity mask that excludes pixels with invalid inverse variance or known artifacts. We use the inverse variance \(\mathrm{ivar}_i\) from the DESI pipeline \citep{guy2023desipipeline} as a reconstruction weight, clipped at 1000 to reduce the influence of extreme weights:
\begin{equation}
    w_i = \min(\mathrm{ivar}_i, 1000).
\end{equation}
The pretraining loss is
\begin{equation}
    \mathcal{L}_{\mathrm{MAE}} =
    \frac{
    \sum_i m_i v_i w_i \left(\hat{f}_i - f_i\right)^2
    }{
    \sum_i m_i v_i w_i
    }.
\end{equation}
After pretraining, we retain only the encoder \(g_s(\cdot)\) and freeze it in all subsequent training stages.

\subsection{Spectral Instruction Tuning}
\label{sec:instruction-tuning}

We perform spectral instruction tuning on DESI-SpecInstruct (Section~\ref{sec:instruction-data}). Each training example consists of a spectrum \(\mathbf{X}_s\), a user question \(\mathbf{q}\), and a target assistant answer \(\mathbf{a}\). During this stage, the SpecMAE encoder and the base parameters of Qwen3-4B are kept frozen. We train only the spectral projector \(P_{\phi}\), the embeddings of the newly added special tokens, and LoRA adapters \citep{hu2022lora} inserted into the linear layers of Qwen3-4B. The LoRA rank is \(r=16\), with scaling factor \(\alpha=32\).

For each example, the SpecMAE encoder produces \(\mathbf{Z}_s=g_s(\mathbf{X}_s)\). The projector maps \(\mathbf{Z}_s\) to \(\mathbf{H}_s\), which is inserted into the language-model input together with the question tokens as defined in Section~\ref{sec:astrospeclm}. The model is optimized with the standard autoregressive language modeling objective, with loss applied only to answer tokens:
\begin{equation}
    \mathcal{L}_{\mathrm{sft}} =
    -\sum_{i=1}^{L}
    \log p_{\theta}(a_i \mid \mathbf{E}_{\mathrm{in}}, a_{<i}).
\end{equation}

\begin{table*}[t]
\centering
\small
\setlength{\tabcolsep}{5pt}
\renewcommand{\arraystretch}{1.12}
\resizebox{\textwidth}{!}{
\begin{tabular}{@{}lccccccc@{}}
\toprule
\multirow{3}{*}{\textbf{Model}}
& \multicolumn{4}{c}{\textbf{Spectrum-grounded Tasks}}
& \multicolumn{3}{c}{\textbf{Language Benchmarks}} \\
\cmidrule(lr){2-5} \cmidrule(lr){6-8}
& \textbf{Object Class}
& \textbf{Stellar Subtype}
& \textbf{Redshift}
& $\boldsymbol{T_\mathrm{eff}}$
& \textbf{MMLU-Redux}
& \textbf{IFEval}
& \textbf{MATH-500} \\
& Acc.
& Acc.
& MAE / R$^2$
& MAE / R$^2$
& Acc.
& Acc.
& Acc. \\
\midrule

SpecMAE     & 98.12 & 85.62 & 0.029/\textbf{0.993} & \textbf{162.3}/\textbf{0.955} & N/A   & N/A   & N/A   \\
AstroCLIP \citep{parker2024astroclip}   & 97.82 & 86.37 & 0.099/0.969 & 330.3/0.942 & N/A   & N/A   & N/A   \\
Qwen3-4B \citep{yang2025qwen3}    & N/A   & N/A   & N/A         & N/A          & \textbf{75.82} & \textbf{69.30} & \textbf{83.60} \\
AstroSpecLM & \textbf{98.56} & \textbf{87.56} & \textbf{0.015}/0.992 & 335.2/0.668 & 74.00 & 65.53 & 80.20 \\
\bottomrule
\end{tabular}}

\caption{Main results on spectrum-grounded tasks and language benchmarks. Best results are shown in bold. N/A indicates tasks that are not applicable due to model input or output constraints.}
\label{tab:main-results}
\end{table*}

\section{Experiments}
\label{sec:experiments}
\begin{figure*}[t]
    \centering
    \includegraphics[width=0.93\linewidth]{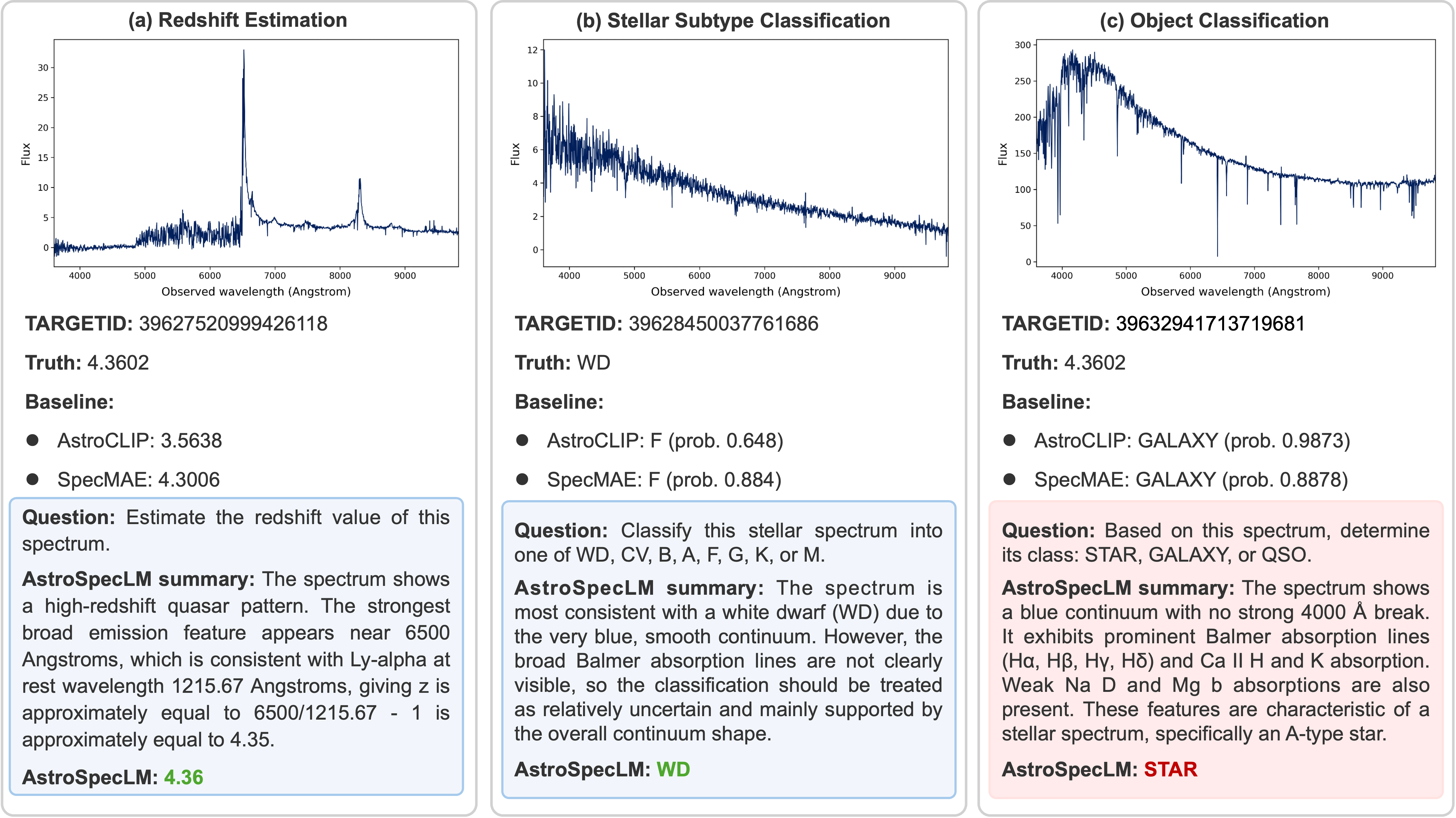}
    \caption{Qualitative examples comparing AstroSpecLM with SpecMAE and AstroCLIP. \textbf{Left:} redshift estimation for a QSO at $z=4.3602$. \textbf{Center:} stellar subtype classification for a white dwarf. \textbf{Right:} a galaxy misclassified as a star by AstroSpecLM.}
    \label{fig:qualitative}
\end{figure*}
\subsection{Experimental Setup}
\label{sec:exp-setup}

\paragraph{Models and Training.}
We first pretrain SpecMAE on \(\mathcal{D}_{\mathrm{enc}}\) and then train AstroSpecLM on DESI-SpecInstruct through spectral instruction tuning. Both stages are conducted on 8 NVIDIA A100-40GB GPUs. SpecMAE pretraining takes approximately 24 hours, and AstroSpecLM instruction tuning takes approximately 3 hours. Detailed optimization hyperparameters for both stages are provided in Appendix~\ref{app:training-details}.

\paragraph{Baselines.}
We compare AstroSpecLM against three baselines. SpecMAE is evaluated by attaching task-specific MLP heads to the frozen spectral encoder used by AstroSpecLM. For AstroCLIP \citep{parker2024astroclip}, we use its pretrained spectral encoder as a frozen feature extractor and train task-specific MLP heads on our training spectra for each classification or regression task. These supervised baselines are not designed for language tasks or free-text generation.
Qwen3-4B \citep{yang2025qwen3} is included as a text-only reference model without additional training, and is evaluated only on the general language benchmarks to indicate the original language capability before spectral instruction tuning.

\paragraph{Evaluation.}
We evaluate AstroSpecLM on the AstroSpecQA benchmark described in Section~\ref{sec:benchmark}. Specifically, we report classification accuracy for both object types and stellar subtypes, alongside Mean Absolute Error (MAE) and $R^2$ coefficients for redshift and effective temperature ($T_{\text{eff}}$) regressions. AstroSpecLM is prompted to first analyze the spectrum and then provide a concise answer after a ``Final Answer'' marker. To assess whether spectral instruction tuning preserves general language capability, we additionally evaluate on three standard benchmarks: MMLU-Redux \citep{gema2024mmluredux} (general knowledge), IFEval \citep{zhou2023ifeval} (instruction following), and MATH-500 \citep{hendrycks2021math,lightman2024verify} (mathematical reasoning).

For classification tasks, we extract the predicted label from the final-answer field using deterministic rules. For numerical tasks, we extract scalar values, round them to at most four decimal places, and compute MAE and \(R^2\) against the catalog target. Outputs that cannot be parsed by rules are passed to DeepSeek-V3.2 for format normalization only, without access to ground-truth labels. We further evaluate explanation faithfulness on all AstroSpecQA outputs using DeepSeek-V3.2 as an LLM judge, which scores whether each explanation is supported by the corresponding held-out spectral facts. The full judging prompt and scoring rubric are provided in Appendix~\ref{app:faithfulness_prompt}. We also conduct a small human meta-evaluation on 100 randomly sampled outputs to assess the reliability of the automatic judge.

\subsection{Main Results}
\label{sec:main-results}

Table~\ref{tab:main-results} reports the results on spectrum-grounded tasks and general language benchmarks.

\paragraph{Spectrum-grounded tasks.}
AstroSpecLM achieves the highest accuracy on both classification tasks, surpassing the specialist supervised baselines on object type (98.56\% vs.\ 98.12\%) and stellar subtype (87.56\% vs.\ 86.37\%). This suggests that the unified generative model can remain competitive with dedicated classifiers while supporting free-form QA.

On redshift estimation, AstroSpecLM attains an MAE of 0.015, lower than SpecMAE's 0.029, while both models achieve comparable R$^2$ scores (0.992 vs. 0.993). This suggests that the generative model can effectively exploit line-position evidence when estimating redshift. Since redshift is closely tied to the observed-to-rest wavelength ratio of identifiable spectral lines, it is well aligned with the model's training objective of producing feature-grounded explanations.

Effective-temperature estimation is more challenging for the generative model. AstroSpecLM's MAE of 335.2\,K and R$^2$ of 0.668 lag substantially behind SpecMAE's 162.3\,K and 0.955. Unlike redshift, effective temperature depends on distributed spectral cues such as continuum shape, line strengths, and molecular-band depths, and is less directly recoverable from a single localized measurement. Autoregressive generation may also introduce additional instability when producing continuous numerical values. Section~\ref{sec:ablation} shows that alignment pretraining can substantially improve this task, suggesting that stronger spectral-language alignment is important for numerical parameter estimation.

\paragraph{Language benchmarks.}
AstroSpecLM retains much of the language capability of the text-only Qwen3-4B baseline after spectral instruction tuning, with MMLU-Redux decreasing from 75.82\% to 74.00\%, IFEval from 69.30\% to 65.53\%, and MATH-500 from 83.60\% to 80.20\%. This 1.8--3.8 point drop indicates moderate degradation rather than severe loss of general language ability, consistent with forgetting effects commonly observed in multimodal fine-tuning \citep{zhai2023catastrophicforgetting,zheng2024mllmforgetting}. Preserving this capability is important for AstroSpecLM, since the model relies on natural-language generation to express evidence-grounded explanations.

\paragraph{Explanation faithfulness.}
We further evaluate the faithfulness of AstroSpecLM's generated explanations on the full AstroSpecQA benchmark using DeepSeek-V3.2 as an LLM judge. AstroSpecLM obtains an average faithfulness score of 4.2/5.0, indicating that most explanations are well supported by the corresponding spectral facts. To calibrate the automatic judge, we manually score 100 randomly sampled outputs using the same rubric and find that the human assessments are generally consistent with the LLM-judge scores, supporting its use as a scalable faithfulness proxy.

\begin{table*}[t]
\centering
\small
\setlength{\tabcolsep}{4pt}
\begin{tabular}{@{}lccccc@{}}
\toprule
\textbf{Model Variant} & \textbf{Object Class} & \textbf{Stellar Subtype} & \textbf{Redshift} & \textbf{$T_{\rm eff}$} & \textbf{MMLU-Redux} \\
& Acc. & Acc. & MAE / R$^2$ & MAE / R$^2$ & Acc. \\
\midrule
\multicolumn{6}{c}{\textit{Training strategy}} \\
\midrule
AstroSpecLM (4B, SFT)                 & \underline{98.56} & \textbf{87.56} & \underline{0.015} / \textbf{0.992} & 335.2 / 0.668 & \textbf{74.00} \\
\quad + Alignment pretraining         & 98.46 & 86.82 & \textbf{0.011} / \textbf{0.992} & \textbf{206.2} / \textbf{0.942} & 70.40 \\
\quad + 2 training epochs             & \textbf{98.78} & 85.62 & 0.031 / \underline{0.975} & 350.6 / 0.712 & 72.10 \\
\midrule
\multicolumn{6}{c}{\textit{Model scale \& token length}} \\
\midrule
AstroSpecLM (1.7B, SFT)               & 88.18 & 71.03 & 0.090 / 0.847 & 374.6 / 0.437 & 64.40 \\
AstroSpecLM (4B, SFT, 260 tok)        & 97.72 & 84.88 & 0.084 / 0.958 & \underline{285.4} / \underline{0.912} & \underline{73.30} \\
\midrule
\multicolumn{6}{c}{\textit{LoRA capacity}} \\
\midrule
AstroSpecLM (4B, SFT, rank=32)        & 98.12 & \underline{87.27} & 0.018 / 0.963 & 327.7 / 0.712 & 73.10 \\
\bottomrule
\end{tabular}
\caption{Ablation results. The first row (4B, SFT, 130 tok, rank=16) is our final configuration. Best results are in bold, second-best results are underlined.}
\label{tab:ablation}
\end{table*}

\subsection{Qualitative Analysis}
\label{sec:qualitative}

To complement the quantitative results, we examine three representative cases that illustrate the strengths and limitations of AstroSpecLM relative to the specialist baselines. Figure~\ref{fig:qualitative} presents the spectra and model outputs for each case.

\paragraph{Case 1: Redshift estimation through line identification.}
The first example (Figure~\ref{fig:qualitative}, left) shows a high-redshift quasar with true $z=4.3602$. AstroSpecLM produces the closest estimate ($z=4.36$) and explicitly identifies Ly$\alpha$ at $\approx 6500$\,\AA\ and C\,IV at $\approx 8305$\,\AA, computing the implied redshifts separately and verifying their mutual consistency. This illustrates how the generative model can outperform direct regression by explicitly modeling the relationship between observed line positions and the redshift formula.

\paragraph{Case 2: Rare subtype classification.}
The second example (Figure~\ref{fig:qualitative}, center) shows a white dwarf (WD) spectrum that both SpecMAE and AstroCLIP misclassify as an F-type star. AstroSpecLM correctly identifies it as WD based on the blue continuum and absence of molecular bands, while explicitly noting that the usual broad Balmer absorption lines are not visible and that the classification is ``relatively uncertain.'' This behavior---providing the correct answer while acknowledging evidence limitations---is a qualitative benefit of the language interface.

\paragraph{Case 3: Misclassification due to feature-level ambiguity.}
The third example (Figure~\ref{fig:qualitative}, right) shows a GALAXY spectrum that both baselines classify correctly. AstroSpecLM misclassifies it as an A-type star, citing prominent Balmer absorption lines and a blue continuum with no strong 4000\,\AA\ break. These features are indeed standard indicators of an A-type stellar atmosphere, and their presence in this galaxy creates genuine ambiguity between an early-type stellar continuum and a galaxy dominated by young stellar populations. The error suggests that the model can over-rely on local absorption patterns when they closely match those of a different class.

\subsection{Ablation Study}
\label{sec:ablation}

We conduct ablation experiments to validate the key design choices in AstroSpecLM, examining whether each component contributes to the model's ability to understand and reason about spectra. Table~\ref{tab:ablation} summarizes the results.

\paragraph{Training strategy.}
We examine two variations of the training pipeline. Adding an intermediate alignment stage --- where the projector and LoRA adapters are first trained on spectrum--caption pairs derived from the same grounded facts before instruction tuning --- substantially improves performance on tasks requiring precise numerical reasoning, particularly effective temperature (R$^2$: 0.668 $\to$ 0.942). However, it also reduces MMLU-Redux from 74.00 to 70.40 and introduces an additional training stage. We therefore report the SFT-only model as our main configuration because it provides a simpler and more balanced trade-off between spectrum-grounded performance and retained language ability. The alignment variant represents a stronger option when numerical accuracy is prioritized. Details of the alignment data are provided in Appendix~\ref{app:alignment}. Extending instruction tuning to two epochs shows signs of overfitting: object classification improves slightly, but both regression tasks degrade, suggesting that longer instruction tuning does not consistently improve spectral grounding.

\paragraph{Model scale and token length.}
Reducing the language model from 4B to 1.7B parameters causes consistent degradation across all tasks, with the sharpest drop in $T_{\rm eff}$ (R$^2$: 0.668 $\to$ 0.437). This suggests that larger language-model capacity is beneficial for integrating spectral representations while preserving instruction-following ability. The spectral token count reflects a fundamental trade-off in how the model attends to the spectrum. With 260 tokens (twice the default), the model improves $T_{\rm eff}$ estimation ($R^2$: 0.668 $\rightarrow$ 0.912), likely because this task benefits from broader continuum shape and distributed line-ratio information. However, redshift accuracy decreases substantially (MAE: 0.015 $\rightarrow$ 0.084), suggesting a trade-off between broad spectral coverage and compact representations for localized line-based prediction. Since this mechanism is not directly verified by attribution analysis, we treat it as a hypothesis and retain 130 tokens as a practical default.

\paragraph{LoRA capacity.}
Increasing the LoRA rank from 16 to 32 produces only small changes across all tasks, confirming that rank-16 provides sufficient adaptation capacity for the model to learn meaningful spectral features without over-parameterization.

\section{Conclusion}
\label{sec:conclusion}
We presented AstroSpecLM, a spectrum-language model that connects 1D DESI spectra with Qwen3-4B for spectrum-grounded question answering. By constructing instruction data through an intermediate fact extraction stage, the model learns to produce answers grounded in observable spectral evidence. Experiments show that the generative approach is viable: AstroSpecLM matches or surpasses specialist supervised baselines on classification and redshift estimation, while effective-temperature regression remains a challenge for token-by-token decoding. In addition, the explanation-faithfulness evaluation indicates that its generated explanations are largely supported by the corresponding spectral facts. These results suggest that 1D scientific measurements can be connected to language models through fact-mediated instruction tuning. Future work may improve numerical grounding through stronger spectral-language alignment, finer wavelength-aware representations, and broader instruction data covering more physical parameters and scientific analysis scenarios.

\bibliography{custom}

\begin{thebibliography}{44}
\expandafter\ifx\csname natexlab\endcsname\relax\def\natexlab#1{#1}\fi

\bibitem[{Alayrac et~al.(2022)Alayrac, Donahue, Luc, Miech, Barr, Hasson, Lenc,
  Mensch, Millican, Reynolds, Ring, Rutherford, Cabi, Han, Gong, Samangooei,
  Monteiro, Menick, Borgeaud, Brock, Nematzadeh, Sharifzadeh, Binkowski,
  Barreira, Vinyals, Zisserman, and Simonyan}]{alayrac2022flamingo}
Jean-Baptiste Alayrac, Jeff Donahue, Pauline Luc, Antoine Miech, Iain Barr,
  Yana Hasson, Karel Lenc, Arthur Mensch, Katie Millican, Malcolm Reynolds,
  Roman Ring, Eliza Rutherford, Serkan Cabi, Tengda Han, Zhitao Gong, Sina
  Samangooei, Marianne Monteiro, Jacob Menick, Sebastian Borgeaud, Andrew
  Brock, Aida Nematzadeh, Sahand Sharifzadeh, Mikolaj Binkowski, Ricardo
  Barreira, Oriol Vinyals, Andrew Zisserman, and Karen Simonyan. 2022.
\newblock \href {https://arxiv.org/abs/2204.14198} {Flamingo: a visual language
  model for few-shot learning}.
\newblock In \emph{Advances in Neural Information Processing Systems},
  volume~35, pages 23716--23736.

\bibitem[{Bai et~al.(2025)Bai, Cai, Chen, Chen, Chen, Cheng, Deng, Ding, Gao,
  Ge et~al.}]{bai2025qwen3vl}
Shuai Bai, Yuxuan Cai, Ruizhe Chen, Keqin Chen, Xionghui Chen, Zesen Cheng,
  Lianghao Deng, Wei Ding, Chang Gao, Chunjiang Ge, et~al. 2025.
\newblock \href {http://arxiv.org/abs/2511.21631} {Qwen3-vl technical report}.
\newblock \emph{arXiv preprint arXiv:2511.21631}.

\bibitem[{Chung et~al.(2024)Chung, Hou, Longpre, Zoph, Tay, Fedus, Li, Wang,
  Dehghani, Brahma, Webson, Gu, Dai, Suzgun, Chen, Chowdhery, Castro-Ros,
  Pellat, Robinson, Valter, Narang, Mishra, Yu, Zhao, Huang, Dai, Yu, Petrov,
  Chi, Dean, Devlin, Roberts, Zhou, Le, and Wei}]{chung2024flan}
Hyung~Won Chung, Le~Hou, Shayne Longpre, Barret Zoph, Yi~Tay, William Fedus,
  Yunxuan Li, Xuezhi Wang, Mostafa Dehghani, Siddhartha Brahma, Albert Webson,
  Shixiang~Shane Gu, Zhuyun Dai, Mirac Suzgun, Xinyun Chen, Aakanksha
  Chowdhery, Alex Castro-Ros, Marie Pellat, Kevin Robinson, Dasha Valter,
  Sharan Narang, Gaurav Mishra, Adams Yu, Vincent Zhao, Yanping Huang, Andrew
  Dai, Hongkun Yu, Slav Petrov, Ed~H. Chi, Jeff Dean, Jacob Devlin, Adam
  Roberts, Denny Zhou, Quoc~V. Le, and Jason Wei. 2024.
\newblock Scaling instruction-finetuned language models.
\newblock \emph{Journal of Machine Learning Research}, 25(70):1--53.

\bibitem[{Dai et~al.(2024)Dai, Lee, Wang, Yang, Liu, Barker, Rintamaki,
  Shoeybi, Catanzaro, and Ping}]{dai2024nvlm}
Wenliang Dai, Nayeon Lee, Boxin Wang, Zhuolin Yang, Zihan Liu, Jon Barker,
  Tuomas Rintamaki, Mohammad Shoeybi, Bryan Catanzaro, and Wei Ping. 2024.
\newblock \href {http://arxiv.org/abs/2409.11402} {Nvlm: Open frontier-class
  multimodal llms}.
\newblock \emph{arXiv preprint arXiv:2409.11402}.

\bibitem[{de~Haan et~al.(2025)de~Haan, Ting, Ghosal, Nguyen, Accomazzi, Wells,
  Ramachandra, Pan, and Sun}]{dehaan2025astrosage}
Tijmen de~Haan, Yuan-Sen Ting, Tirthankar Ghosal, Tuan~Dung Nguyen, Alberto
  Accomazzi, Azton Wells, Nesar Ramachandra, Rui Pan, and Zechang Sun. 2025.
\newblock \href {https://doi.org/10.1038/s41598-025-97131-y} {Achieving gpt-4o
  level performance in astronomy with a specialized 8b-parameter large language
  model}.
\newblock \emph{Scientific Reports}.

\bibitem[{{DeepSeek-AI}(2025)}]{deepseek2025v32}
{DeepSeek-AI}. 2025.
\newblock \href {https://arxiv.org/abs/2512.02556} {{DeepSeek-V3.2}: Pushing
  the frontier of open large language models}.
\newblock \emph{arXiv preprint arXiv:2512.02556}.

\bibitem[{{DESI Collaboration} et~al.(2026){DESI Collaboration}, Abdul~Karim,
  Adame, Aguado, Aguilar et~al.}]{desi2026dr1}
{DESI Collaboration}, M.~Abdul~Karim, A.~G. Adame, D.~Aguado, J.~Aguilar,
  et~al. 2026.
\newblock \href {https://doi.org/10.3847/1538-3881/ae4c43} {Data release 1 of
  the dark energy spectroscopic instrument}.
\newblock \emph{The Astronomical Journal}, 171(5):285.

\bibitem[{{DESI Collaboration} et~al.(2024){DESI Collaboration}, Adame,
  Aguilar, Ahlen, Alam, Aldering, Alexander, Alfarsy, Allende~Prieto
  et~al.}]{desicollaboration2024edr}
{DESI Collaboration}, A.~G. Adame, J.~Aguilar, S.~Ahlen, S.~Alam, G.~Aldering,
  D.~M. Alexander, R.~Alfarsy, C.~Allende~Prieto, et~al. 2024.
\newblock \href {https://doi.org/10.3847/1538-3881/ad3217} {The early data
  release of the dark energy spectroscopic instrument}.
\newblock \emph{The Astronomical Journal}, 168(2):58.

\bibitem[{Ding et~al.(2023)Ding, Chen, Xu, Qin, Hu, Liu, Sun, and
  Zhou}]{ding2023ultrachat}
Ning Ding, Yulin Chen, Bokai Xu, Yujia Qin, Shengding Hu, Zhiyuan Liu, Maosong
  Sun, and Bowen Zhou. 2023.
\newblock \href {https://doi.org/10.18653/v1/2023.emnlp-main.183} {Enhancing
  chat language models by scaling high-quality instructional conversations}.
\newblock In \emph{Proceedings of the 2023 Conference on Empirical Methods in
  Natural Language Processing}, pages 3029--3051, Singapore. Association for
  Computational Linguistics.

\bibitem[{Gema et~al.(2025)Gema, Leang, Hong, Devoto, Mancino, Saxena, He,
  Zhao, Du, Madani, Barale, McHardy, Harris, Kaddour, van Krieken, and
  Minervini}]{gema2024mmluredux}
Aryo~Pradipta Gema, Joshua Ong~Jun Leang, Giwon Hong, Alessio Devoto, Alberto
  Carlo~Maria Mancino, Rohit Saxena, Xuanli He, Yu~Zhao, Xiaotang Du, Mohammad
  Reza~Ghasemi Madani, Claire Barale, Robert McHardy, Joshua Harris, Jean
  Kaddour, Emile van Krieken, and Pasquale Minervini. 2025.
\newblock \href {https://doi.org/10.18653/V1/2025.NAACL-LONG.262} {Are we done
  with mmlu?}
\newblock In \emph{Proceedings of the 2025 Conference of the Nations of the
  Americas Chapter of the Association for Computational Linguistics: Human
  Language Technologies, {NAACL} 2025 - Volume 1: Long Papers, Albuquerque, New
  Mexico, USA, April 29 - May 4, 2025}, pages 5069--5096. Association for
  Computational Linguistics.

\bibitem[{Grattafiori et~al.(2024)Grattafiori, Dubey, Jauhri, Pandey, Kadian,
  Al-Dahle, Letman, Mathur, Schelten, Vaughan et~al.}]{grattafiori2024llama3}
Aaron Grattafiori, Abhimanyu Dubey, Abhinav Jauhri, Abhinav Pandey, Abhishek
  Kadian, Ahmad Al-Dahle, Aiesha Letman, Akhil Mathur, Alan Schelten, Alex
  Vaughan, et~al. 2024.
\newblock \href {http://arxiv.org/abs/2407.21783} {The llama 3 herd of models}.
\newblock \emph{arXiv preprint arXiv:2407.21783}.

\bibitem[{Guy et~al.(2023)Guy, Bailey, Kremin, Alam, Alexander
  et~al.}]{guy2023desipipeline}
J.~Guy, S.~Bailey, A.~Kremin, S.~Alam, D.~M. Alexander, et~al. 2023.
\newblock \href {https://doi.org/10.3847/1538-3881/acb212} {The spectroscopic
  data processing pipeline for the dark energy spectroscopic instrument}.
\newblock \emph{The Astronomical Journal}, 165(4):144.

\bibitem[{He et~al.(2022)He, Chen, Xie, Li, Doll\'ar, and Girshick}]{he2022mae}
Kaiming He, Xinlei Chen, Saining Xie, Yanghao Li, Piotr Doll\'ar, and Ross
  Girshick. 2022.
\newblock \href
  {https://openaccess.thecvf.com/content/CVPR2022/html/He_Masked_Autoencoders_Are_Scalable_Vision_Learners_CVPR_2022_paper.html}
  {Masked autoencoders are scalable vision learners}.
\newblock In \emph{Proceedings of the IEEE/CVF Conference on Computer Vision
  and Pattern Recognition (CVPR)}, pages 16000--16009.

\bibitem[{Hendrycks et~al.(2021)Hendrycks, Burns, Kadavath, Arora, Basart,
  Tang, Song, and Steinhardt}]{hendrycks2021math}
Dan Hendrycks, Collin Burns, Saurav Kadavath, Akul Arora, Steven Basart, Eric
  Tang, Dawn Song, and Jacob Steinhardt. 2021.
\newblock \href {https://arxiv.org/abs/2103.03874} {Measuring mathematical
  problem solving with the {MATH} dataset}.
\newblock In \emph{Advances in Neural Information Processing Systems},
  volume~34, pages 7949--7962.

\bibitem[{Hu et~al.(2025)Hu, Xu, Zhang, Ye, Yan, Zhang, Jin, Huang, and
  Zhou}]{hu2025docowl2}
Anwen Hu, Haiyang Xu, Liang Zhang, Jiabo Ye, Ming Yan, Ji~Zhang, Qin Jin, Fei
  Huang, and Jingren Zhou. 2025.
\newblock \href {https://doi.org/10.18653/v1/2025.acl-long.291}
  {m{PLUG}-{D}oc{O}wl2: High-resolution compressing for {OCR}-free multi-page
  document understanding}.
\newblock In \emph{Proceedings of the 63rd Annual Meeting of the Association
  for Computational Linguistics (Volume 1: Long Papers)}, pages 5817--5834,
  Vienna, Austria. Association for Computational Linguistics.

\bibitem[{Hu et~al.(2022)Hu, Shen, Wallis, Allen-Zhu, Li, Wang, Wang, and
  Chen}]{hu2022lora}
Edward~J. Hu, Yelong Shen, Phillip Wallis, Zeyuan Allen-Zhu, Yuanzhi Li, Shean
  Wang, Lu~Wang, and Weizhu Chen. 2022.
\newblock \href {https://openreview.net/forum?id=nZeVKeeFYf9} {{LoRA}: Low-rank
  adaptation of large language models}.
\newblock In \emph{International Conference on Learning Representations}.

\bibitem[{Islam and Fox(2026)}]{islam2026omnispectra}
Md~Khairul Islam and Judy Fox. 2026.
\newblock \href {https://doi.org/10.48550/arXiv.2601.15351} {Omnispectra: A
  unified foundation model for native resolution astronomical spectra}.
\newblock \emph{arXiv preprint arXiv:2601.15351}.

\bibitem[{Jia et~al.(2026)Jia, Zhang, Luo, Li, Ye, Lu, Hou, and
  Zhao}]{jia2026speco3}
Minghui Jia, Qichao Zhang, Ali Luo, Linjing Li, Shuo Ye, Hailing Lu, Wen Hou,
  and Dongbin Zhao. 2026.
\newblock \href {http://arxiv.org/abs/2601.06498} {Spec-o3: A tool-augmented
  vision-language agent for rare celestial object candidate vetting via
  automated spectral inspection}.

\bibitem[{Koblischke and Bovy(2024)}]{koblischke2024spectrafm}
Nolan Koblischke and Jo~Bovy. 2024.
\newblock \href {http://arxiv.org/abs/2411.04750} {Spectrafm: Tuning into
  stellar foundation models}.
\newblock \emph{arXiv preprint arXiv:2411.04750}.

\bibitem[{Li et~al.(2024)Li, Zhang, Guo, Zhang, Li, Zhang, Zhang, Zhang, Li,
  Liu, and Li}]{li2024llavaonevision}
Bo~Li, Yuanhan Zhang, Dong Guo, Renrui Zhang, Feng Li, Hao Zhang, Kaichen
  Zhang, Peiyuan Zhang, Yanwei Li, Ziwei Liu, and Chunyuan Li. 2024.
\newblock \href {http://arxiv.org/abs/2408.03326} {Llava-onevision: Easy visual
  task transfer}.
\newblock \emph{arXiv preprint arXiv:2408.03326}.

\bibitem[{Li et~al.(2023)Li, Li, Savarese, and Hoi}]{li2023blip2}
Junnan Li, Dongxu Li, Silvio Savarese, and Steven Hoi. 2023.
\newblock \href {https://proceedings.mlr.press/v202/li23q.html} {{BLIP}-2:
  Bootstrapping language-image pre-training with frozen image encoders and
  large language models}.
\newblock In \emph{Proceedings of the 40th International Conference on Machine
  Learning}, volume 202 of \emph{Proceedings of Machine Learning Research},
  pages 19730--19742. PMLR.

\bibitem[{Lightman et~al.(2024)Lightman, Kosaraju, Burda, Edwards, Baker, Lee,
  Leike, Schulman, Sutskever, and Cobbe}]{lightman2024verify}
Hunter Lightman, Vineet Kosaraju, Yura Burda, Harri Edwards, Bowen Baker, Teddy
  Lee, Jan Leike, John Schulman, Ilya Sutskever, and Karl Cobbe. 2024.
\newblock \href {https://openreview.net/forum?id=v8L0pN6EOi} {Let's verify step
  by step}.
\newblock In \emph{International Conference on Learning Representations}.

\bibitem[{Lin et~al.(2024)Lin, Ye, Zhu, Cui, Ning, Jin, and
  Yuan}]{lin2024videollava}
Bin Lin, Yang Ye, Bin Zhu, Jiaxi Cui, Munan Ning, Peng Jin, and Li~Yuan. 2024.
\newblock \href {https://doi.org/10.18653/v1/2024.emnlp-main.342}
  {Video-{LL}a{VA}: Learning united visual representation by alignment before
  projection}.
\newblock In \emph{Proceedings of the 2024 Conference on Empirical Methods in
  Natural Language Processing}, pages 5971--5984, Miami, Florida, USA.
  Association for Computational Linguistics.

\bibitem[{Liu et~al.(2023)Liu, Li, Wu, and Lee}]{liu2023llava}
Haotian Liu, Chunyuan Li, Qingyang Wu, and Yong~Jae Lee. 2023.
\newblock \href
  {https://papers.nips.cc/paper_files/paper/2023/hash/6dcf277ea32ce3288914faf369fe6de0-Abstract-Conference.html}
  {Visual instruction tuning}.
\newblock In \emph{Advances in Neural Information Processing Systems},
  volume~36.

\bibitem[{Maaz et~al.(2024)Maaz, Rasheed, Khan, and
  Khan}]{maaz2024videochatgpt}
Muhammad Maaz, Hanoona Rasheed, Salman Khan, and Fahad Khan. 2024.
\newblock \href {https://doi.org/10.18653/v1/2024.acl-long.679}
  {Video-{C}hat{GPT}: Towards detailed video understanding via large vision and
  language models}.
\newblock In \emph{Proceedings of the 62nd Annual Meeting of the Association
  for Computational Linguistics (Volume 1: Long Papers)}, pages 12585--12602,
  Bangkok, Thailand. Association for Computational Linguistics.

\bibitem[{Mishra-Sharma et~al.(2024)Mishra-Sharma, Song, and
  Thaler}]{mishrasharma2024paperclip}
Siddharth Mishra-Sharma, Yiding Song, and Jesse Thaler. 2024.
\newblock \href {http://arxiv.org/abs/2403.08851} {Paperclip: Associating
  astronomical observations and natural language with multi-modal models}.

\bibitem[{Nguyen et~al.(2023)Nguyen, Ting, Ciuca, O{'}Neill, Sun,
  Jab{\l}o{\'n}ska, Kruk, Perkowski, Miller, Li, Peek, Iyer, Rozanski,
  Khetarpal, Zaman, Brodrick, Rodriguez~Mendez, Bui, Goodman, Accomazzi,
  Naiman, Cranney, Schawinski, and Raileanu}]{nguyen2023astrollama}
Tuan~Dung Nguyen, Yuan-Sen Ting, Ioana Ciuca, Charles O{'}Neill, Ze-Chang Sun,
  Maja Jab{\l}o{\'n}ska, Sandor Kruk, Ernest Perkowski, Jack Miller, Jason
  Jason~Jingsh Li, Josh Peek, Kartheik Iyer, Tomasz Rozanski, Pranav Khetarpal,
  Sharaf Zaman, David Brodrick, Sergio~J. Rodriguez~Mendez, Thang Bui, Alyssa
  Goodman, Alberto Accomazzi, Jill Naiman, Jesse Cranney, Kevin Schawinski, and
  Roberta Raileanu. 2023.
\newblock \href {https://doi.org/10.18653/v1/2023.wiesp-1.7} {{A}stro{LL}a{MA}:
  Towards specialized foundation models in astronomy}.
\newblock In \emph{Proceedings of the Second Workshop on Information Extraction
  from Scientific Publications}, pages 49--55, Bali, Indonesia. Association for
  Computational Linguistics.

\bibitem[{Ouyang et~al.(2022)Ouyang, Wu, Jiang, Almeida, Wainwright, Mishkin,
  Zhang, Agarwal, Slama, Ray, Schulman, Hilton, Kelton, Miller, Simens, Askell,
  Welinder, Christiano, Leike, and Lowe}]{ouyang2022instructgpt}
Long Ouyang, Jeff Wu, Xu~Jiang, Diogo Almeida, Carroll~L. Wainwright, Pamela
  Mishkin, Chong Zhang, Sandhini Agarwal, Katarina Slama, Alex Ray, John
  Schulman, Jacob Hilton, Fraser Kelton, Luke Miller, Maddie Simens, Amanda
  Askell, Peter Welinder, Paul Christiano, Jan Leike, and Ryan Lowe. 2022.
\newblock Training language models to follow instructions with human feedback.
\newblock In \emph{Advances in Neural Information Processing Systems},
  volume~35, pages 27730--27744.

\bibitem[{Parker et~al.(2024)Parker, Lanusse, Golkar, Bietti, Cranmer,
  Eickenberg, Krawezik, McCabe, Ohana, Pettee, Regaldo-Saint~Blancard, Cho, and
  Ho}]{parker2024astroclip}
Liam Parker, Francois Lanusse, Siavash Golkar, Alberto Bietti, Miles Cranmer,
  Michael Eickenberg, Geraud Krawezik, Michael McCabe, Ruben Ohana, Mariel
  Pettee, Bruno Regaldo-Saint~Blancard, Kyunghyun Cho, and Shirley Ho. 2024.
\newblock \href {https://doi.org/10.1093/mnras/stae1450} {Astroclip: A
  cross-modal foundation model for galaxies}.
\newblock \emph{Monthly Notices of the Royal Astronomical Society},
  531(4):4990--5011.

\bibitem[{Parker et~al.(2025)Parker, Lanusse, Shen, Liu, Hehir, Sarra, Meyer,
  Bowles, Wagner-Carena, Qu, Golkar, Bietti, Bourfoune, Cornette, Hirashima,
  Krawezik, Ohana, Lourie, McCabe, Morel, Mukhopadhyay, Pettee, Cho, Cranmer,
  and Ho}]{parker2025aion}
Liam Parker, Francois Lanusse, Jeff Shen, Ollie Liu, Tom Hehir, Leopoldo Sarra,
  Lucas Meyer, Micah Bowles, Sebastian Wagner-Carena, Helen Qu, Siavash Golkar,
  Alberto Bietti, Hatim Bourfoune, Pierre Cornette, Keiya Hirashima, Geraud
  Krawezik, Ruben Ohana, Nicholas Lourie, Michael McCabe, Rudy Morel, Payel
  Mukhopadhyay, Mariel Pettee, Kyunghyun Cho, Miles Cranmer, and Shirley Ho.
  2025.
\newblock \href
  {https://proceedings.neurips.cc/paper_files/paper/2025/file/893df77404832e974b097b361ef49623-Paper-Conference.pdf}
  {Aion-1: Omnimodal foundation model for astronomical sciences}.
\newblock In \emph{Advances in Neural Information Processing Systems},
  volume~38, pages 95386--95428. Curran Associates, Inc.

\bibitem[{Ramachandra et~al.(2025)Ramachandra, Ting, Sun, Wells, and
  Habib}]{ramachandra2025speak}
Nesar Ramachandra, Yuan-Sen Ting, Zechang Sun, Azton Wells, and Salman Habib.
  2025.
\newblock \href {http://arxiv.org/abs/2508.10075} {Teaching llms to speak
  spectroscopy}.
\newblock \emph{arXiv preprint arXiv:2508.10075}.

\bibitem[{Riggi et~al.(2025)Riggi, Cecconello, Pilzer, Palazzo, Gupta, Hopkins,
  Trigilio, and Umana}]{riggi2025radiollava}
S.~Riggi, T.~Cecconello, A.~Pilzer, S.~Palazzo, N.~Gupta, A.~M. Hopkins,
  C.~Trigilio, and G.~Umana. 2025.
\newblock \href {http://arxiv.org/abs/2503.23859} {Evaluating small
  vision-language models as ai assistants for radio astronomical source
  analysis tasks}.
\newblock \emph{arXiv preprint arXiv:2503.23859}.

\bibitem[{Taylor et~al.(2022)Taylor, Kardas, Cucurull, Scialom, Hartshorn,
  Saravia, Poulton, Kerkez, and Stojnic}]{taylor2022galactica}
Ross Taylor, Marcin Kardas, Guillem Cucurull, Thomas Scialom, Anthony
  Hartshorn, Elvis Saravia, Andrew Poulton, Viktor Kerkez, and Robert Stojnic.
  2022.
\newblock \href {https://doi.org/10.48550/arXiv.2211.09085} {Galactica: A large
  language model for science}.
\newblock \emph{arXiv preprint arXiv:2211.09085}.

\bibitem[{Vaswani et~al.(2017)Vaswani, Shazeer, Parmar, Uszkoreit, Jones,
  Gomez, Kaiser, and Polosukhin}]{vaswani2017attention}
Ashish Vaswani, Noam Shazeer, Niki Parmar, Jakob Uszkoreit, Llion Jones,
  Aidan~N. Gomez, {\L}ukasz Kaiser, and Illia Polosukhin. 2017.
\newblock \href {https://papers.nips.cc/paper/7181-attention-is-all-you-need}
  {Attention is all you need}.
\newblock In \emph{Advances in Neural Information Processing Systems},
  volume~30.

\bibitem[{Wang et~al.(2023)Wang, Kordi, Mishra, Liu, Smith, Khashabi, and
  Hajishirzi}]{wang2023selfinstruct}
Yizhong Wang, Yeganeh Kordi, Swaroop Mishra, Alisa Liu, Noah~A. Smith, Daniel
  Khashabi, and Hannaneh Hajishirzi. 2023.
\newblock \href {https://doi.org/10.18653/v1/2023.acl-long.754} {Self-instruct:
  Aligning language models with self-generated instructions}.
\newblock In \emph{Proceedings of the 61st Annual Meeting of the Association
  for Computational Linguistics (Volume 1: Long Papers)}, pages 13484--13508,
  Toronto, Canada. Association for Computational Linguistics.

\bibitem[{Yang et~al.(2025)Yang, Li, Yang, Zhang, Hui, Zheng, Yu, Gao, Huang,
  Lv et~al.}]{yang2025qwen3}
An~Yang, Anfeng Li, Baosong Yang, Beichen Zhang, Binyuan Hui, Bo~Zheng, Bowen
  Yu, Chang Gao, Chengen Huang, Chenxu Lv, et~al. 2025.
\newblock \href {http://arxiv.org/abs/2505.09388} {Qwen3 technical report}.

\bibitem[{Zaman et~al.(2025)Zaman, Smith, Khetarpal, Chakrabarty, Ginolfi,
  Huertas-Company, Jab{\l}o{\'n}ska, Kruk, Le~Lain, Rodr{\'i}guez~M{\'e}ndez,
  and Tanoglidis}]{zaman2025astrollava}
Sharaf Zaman, Michael~J. Smith, Pranav Khetarpal, Rishabh Chakrabarty, Michele
  Ginolfi, Marc Huertas-Company, Maja Jab{\l}o{\'n}ska, Sandor Kruk, Matthieu
  Le~Lain, Sergio~Jos{\'e} Rodr{\'i}guez~M{\'e}ndez, and Dimitrios Tanoglidis.
  2025.
\newblock \href {http://arxiv.org/abs/2504.08583} {Astrollava: Towards the
  unification of astronomical data and natural language}.

\bibitem[{Zhai et~al.(2023)Zhai, Tong, Li, Cai, Qu, Lee, and
  Ma}]{zhai2023catastrophicforgetting}
Yuexiang Zhai, Shengbang Tong, Xiao Li, Mu~Cai, Qing Qu, Yong~Jae Lee, and
  Yi~Ma. 2023.
\newblock \href {https://arxiv.org/abs/2309.10313} {Investigating the
  catastrophic forgetting in multimodal large language models}.
\newblock \emph{arXiv preprint arXiv:2309.10313}.

\bibitem[{Zhang et~al.(2024)Zhang, Chen, Jin, Wang, Ji, Wang, and
  Han}]{zhang2024scientificllmsurvey}
Yu~Zhang, Xiusi Chen, Bowen Jin, Sheng Wang, Shuiwang Ji, Wei Wang, and Jiawei
  Han. 2024.
\newblock \href {https://doi.org/10.18653/V1/2024.EMNLP-MAIN.498} {A
  comprehensive survey of scientific large language models and their
  applications in scientific discovery}.
\newblock In \emph{Proceedings of the 2024 Conference on Empirical Methods in
  Natural Language Processing, {EMNLP} 2024, Miami, FL, USA, November 12-16,
  2024}, pages 8783--8817. Association for Computational Linguistics.

\bibitem[{Zheng et~al.(2024)Zheng, Ma, Liu, Wu, and
  Feng}]{zheng2024mllmforgetting}
Junhao Zheng, Qianli Ma, Zhen Liu, Binquan Wu, and Huawen Feng. 2024.
\newblock \href {https://arxiv.org/abs/2401.09181} {Beyond anti-forgetting:
  Multimodal continual instruction tuning with positive forward transfer}.
\newblock \emph{arXiv preprint arXiv:2401.09181}.

\bibitem[{Zhong et~al.(2024)Zhong, Napolitano, Heneka, Li, Bauer
  et~al.}]{zhong2024gasnet}
Fucheng Zhong, Nicola~R. Napolitano, Caroline Heneka, Rui Li, Franz~Erik Bauer,
  et~al. 2024.
\newblock \href {https://doi.org/10.1093/mnras/stae1461} {Galaxy spectra neural
  network ({GaSNet}). {II}. using deep learning for spectral classification and
  redshift predictions}.
\newblock \emph{Monthly Notices of the Royal Astronomical Society},
  532(1):643--665.

\bibitem[{Zhou et~al.(2023)Zhou, Lu, Mishra, Brahma, Basu, Luan, Zhou, and
  Hou}]{zhou2023ifeval}
Jeffrey Zhou, Tianjian Lu, Swaroop Mishra, Siddhartha Brahma, Sujoy Basu,
  Yi~Luan, Denny Zhou, and Le~Hou. 2023.
\newblock \href {https://arxiv.org/abs/2311.07911} {Instruction-following
  evaluation for large language models}.
\newblock \emph{arXiv preprint arXiv:2311.07911}.

\bibitem[{Zhu et~al.(2024)Zhu, Chen, Shen, Li, and Elhoseiny}]{zhu2024minigpt4}
Deyao Zhu, Jun Chen, Xiaoqian Shen, Xiang Li, and Mohamed Elhoseiny. 2024.
\newblock \href {https://openreview.net/forum?id=1tZbq88f27} {{MiniGPT}-4:
  Enhancing vision-language understanding with advanced large language models}.
\newblock In \emph{International Conference on Learning Representations}.

\bibitem[{Zhu et~al.(2025)Zhu, Wang, Chen, Liu, Ye, Gu, Duan, Tian, Su, Shao
  et~al.}]{zhu2025internvl3}
Jinguo Zhu, Weiyun Wang, Zhe Chen, Zhaoyang Liu, Shenglong Ye, Lixin Gu, Yuchen
  Duan, Hao Tian, Weijie Su, Jie Shao, et~al. 2025.
\newblock \href {http://arxiv.org/abs/2504.10479} {Internvl3: Exploring
  advanced training and test-time recipes for open-source multimodal models}.
\newblock \emph{arXiv preprint arXiv:2504.10479}.

\end{thebibliography}

\onecolumn
\appendix
\section*{Supplementary Material}
\small
\setlength{\parskip}{2pt}
\setlength{\textfloatsep}{10pt}
\setlength{\intextsep}{10pt}
\captionsetup{font=small,skip=5pt}
\tcbset{before skip=8pt,after skip=5pt}
\label{sec:appendix}

\section{Data Details}
\label{app:data-details}

\subsection{Spectrum Corpus and Preprocessing}
\label{app:corpus-details}
\paragraph{Background on DESI.}
The Dark Energy Spectroscopic Instrument (DESI) is a large-scale multi-object spectroscopic survey designed to obtain optical spectra for millions of astronomical sources, including galaxies, quasars, and stars. DESI operates on the Mayall 4-meter telescope and uses thousands of fibers to collect spectra over a wide field of view, with its spectrographs covering the optical wavelength range of approximately \(3600\)--\(9800\,\text{\AA}\). Its public data releases provide calibrated 1D spectra together with pipeline-derived annotations such as object class, redshift, and quality indicators \citep{desi2026dr1}. These properties make DESI particularly suitable for our setting: the spectra are large-scale, homogeneous, and physically interpretable, while the accompanying catalog information enables the construction of grounded supervision for spectrum-language modeling.

We use DESI DR1 1D spectra. Each spectrum provides a flux sequence and an inverse-variance sequence defined on a common wavelength grid of 7781 pixels (3600--9824\,\AA, step 0.8\,\AA). For encoder pretraining, we use 2,853,354 spectra balanced across STAR (951,118), GALAXY (951,118), and QSO (951,118), split 95/5 into training and validation. All pretraining spectra satisfy $\texttt{ZWARN}=0$, finite redshift uncertainty, and $\text{SNR}>2$.

For instruction-data construction, we sample a higher-quality subset of 58,558 spectra from the same collection, requiring $\texttt{ZWARN}=0$, $\texttt{COADD\_FIBERSTATUS}=0$, finite $\texttt{ZERR}$ and $\texttt{DELTACHI2}$, $0 \leq \texttt{ZERR} < 0.005$, $\texttt{DELTACHI2}>30$, and predominantly $\text{SNR}>5$ (with a small fraction of lower-SNR spectra retained for robustness).

Each flux sequence is normalized to zero mean and unit variance before being fed to the spectrum encoder. The inverse variance is used only for weighting the masked reconstruction loss and for quality filtering, not as encoder input.

\paragraph{Data license.}
DESI DR1 is publicly available under the Creative Commons Attribution 4.0 International License. We use the spectra and catalog annotations in accordance with the DESI data-access terms and citation requirements.

\subsection{Grounded Fact Extraction}
\label{app:fact-details}

For each spectrum selected for language-data construction, we extract a compact set of \emph{grounded facts} that combine catalog annotations with quantities measured directly from the spectrum. Table~\ref{tab:app-fact-schema} summarizes the fact types.

Spectral lines are measured in the rest frame using local continuum subtraction. For each predefined line, we integrate positive (emission) or negative (absorption) residuals within a line-centered window, using continuum sidebands for local baseline estimation. Lines are marked invalid if they fall in regions of insufficient wavelength coverage, have too few valid pixels, or yield physically implausible equivalent widths. The set of targeted lines depends on the object type; Table~\ref{tab:app-spectral-evidence} lists the representative features used for each class. K and M stars additionally include molecular-band indices computed from flux ratios between feature and continuum bands.

\begin{table}[H]
\centering
\small
\begin{tabular}{lll}
\toprule
\textbf{Fact Type} & \textbf{Fields} & \textbf{Source} \\
\midrule
Object labels & spectral type, subtype, redshift & catalog \\
Continuum & slope label, red-to-blue ratio & measured \\
D\textsubscript{n}4000 & value, strength label & measured \\
Spectral lines & rest/observed wavelength, EW, significance, status & measured \\
Molecular bands & CaH, TiO, VO indices (K/M stars) & measured \\
Stellar parameters & effective temperature (when available) & catalog \\
\bottomrule
\end{tabular}
\caption{Structured fact types used as grounding references.}
\label{tab:app-fact-schema}
\end{table}

\begin{table}[H]
\centering
\small
\begin{tabularx}{\textwidth}{@{}lX@{}}
\toprule
\textbf{Object Class} & \textbf{Representative Spectral Features} \\
\midrule
GALAXY & [O\,II], [O\,III], H$\alpha$, H$\beta$, [N\,II], [S\,II], Ca H/K, Mg b, Na D \\
QSO & Ly$\alpha$, N\,V, Si\,IV+O\,IV], C\,IV, C\,III], Mg\,II, H$\alpha$, H$\beta$, [O\,III] \\
STAR & Ca H/K, Balmer lines, Mg b, Na D, H$\alpha$, Ca II triplet \\
K/M STAR & CaH, TiO molecular bands \\
\bottomrule
\end{tabularx}
\caption{Class-specific spectral lines and bands used in fact extraction.}
\label{tab:app-spectral-evidence}
\end{table}

\clearpage
\subsection{Instruction Generation}
\label{app:prompt-details}
We use DeepSeek-V3.2 to generate instruction-following conversations conditioned on the grounded facts. The full generation prompt is shown in Figure~\ref{fig:instruction-generation-prompt}.

\begin{figure}[H]
\centering
\begin{promptbox}
\footnotesize
Act like an expert with extensive experience writing in the field of astrophysics.

\medskip
\textbf{Objective:} \\
You are an AI data generator for instruction tuning of a spectral encoder + language model.

The final model will NOT see the text description below. During training, spectrum placeholders are inserted by the data collator, so the JSON output must contain natural user questions only. Do not include the literal token \texttt{<spectrum>}.

You are given a hidden reference description of one astronomical spectrum. Use it only as grounding information to create instruction conversations.

Your task is to generate instruction-tuning conversations for a model that answers questions about the spectrum.

\medskip
\textbf{Important rules:}
\begin{enumerate}[leftmargin=*,nosep]
\item Do NOT mention the hidden reference description, summary, JSON, fields, annotations, facts, labels, or provided text.
\item The user questions must sound like natural questions asked while viewing a spectrum.
\item The assistant answers must sound as if the model is analyzing the spectrum directly.
\item Use only information supported by the hidden reference description.
\item Do not invent spectral lines, parameters, object types, wavelengths, or physical interpretations not present in the hidden reference.
\item It is acceptable to use approximate physical language such as ``around 3800\,K'' or ``consistent with an M-type dwarf''.
\item Avoid questions such as ``What does the summary say?'', ``According to the provided data\dots'', and questions about catalog identifiers, dataset metadata, or annotation fields.
\item Prefer questions such as ``What type of object does this spectrum suggest?'', ``What features dominate this spectrum?'', ``What can be inferred from the continuum shape?'', and ``Why does this look like an M-type star?''
\item Generate samples that may include either or both of the following types:
\begin{itemize}[leftmargin=*,nosep]
    \item single-turn simple QA samples or reasoning QA samples
    \item multi-turn conversational samples with follow-up questions that dive deeper into the spectrum
\end{itemize}
\end{enumerate}

\medskip
\textbf{Output format:}

The output format should follow this style:

\begin{verbatim}
[
  {"type": "single_turn", "messages": [
    {"role": "user", "content": ""},
    {"role": "assistant", "content": ""}
  ]},
  {"type": "multi_turn", "messages": [
    {"role": "user", "content": ""},
    {"role": "assistant", "content": ""},
    {"role": "user", "content": ""},
    {"role": "assistant", "content": ""}
  ]}
]
\end{verbatim}

Multi-turn samples can contain user/assistant conversations of any number of turns.

Generate diverse questions covering:
\begin{itemize}[leftmargin=*,nosep]
  \item object type or spectral subtype
  \item redshift
  \item continuum shape
  \item dominant spectral features
  \item molecular absorption bands
  \item stellar temperature
  \item physical interpretation of the spectrum
  \item and other spectrum-related questions that are supported by the hidden reference
\end{itemize}

Return valid JSON only. Do not include markdown. Return a JSON array, not an object.
\end{promptbox}
\caption{Full prompt used for instruction data generation.}
\label{fig:instruction-generation-prompt}
\end{figure}

\clearpage
\subsection{AstroSpecQA Benchmark Details}
\label{app:benchmark}
Table~\ref{tab:astrospecqa_object} reports the object-class distribution for the object-classification split. The three major classes are approximately balanced, reducing the risk that object-level accuracy is dominated by a single class. Table~\ref{tab:astrospecqa_stellar} reports the stellar-subtype distribution. This split is naturally imbalanced: common stellar types such as K, G, M, and F dominate the evaluation set, while rare categories such as B stars and cataclysmic variables contain very few examples.

\begin{table}[H]
\centering
\begin{minipage}[t]{0.47\textwidth}
\vspace{0pt}
\centering
\small
\begin{tabular}{lrr}
\toprule
\textbf{Object Class} & \textbf{Count} & \textbf{Ratio} \\
\midrule
STAR & 1,753 & 35.06\% \\
GALAXY & 1,648 & 32.96\% \\
QSO & 1,599 & 31.98\% \\
\bottomrule
\end{tabular}
\caption{Object-class distribution in the AstroSpecQA object-classification split.}
\label{tab:astrospecqa_object}
\end{minipage}\hfill
\begin{minipage}[t]{0.47\textwidth}
\vspace{0pt}
\centering
\small
\begin{tabular}{lrr}
\toprule
\textbf{Stellar Subtype} & \textbf{Count} & \textbf{Ratio} \\
\midrule
K  & 620 & 35.37\% \\
G  & 386 & 22.02\% \\
M  & 322 & 18.37\% \\
F  & 274 & 15.63\% \\
WD & 94  & 5.36\% \\
A  & 53  & 3.02\% \\
B  & 3   & 0.17\% \\
CV & 1   & 0.06\% \\
\bottomrule
\end{tabular}
\caption{Stellar-subtype distribution in the AstroSpecQA stellar-subtype split.}
\label{tab:astrospecqa_stellar}
\end{minipage}
\end{table}

\subsection{Alignment Data}
\label{app:alignment}

For the alignment pretraining experiment (Section~\ref{sec:ablation}), we construct 474,051 spectrum--caption pairs from the same grounded facts used for instruction data generation. Each caption is a concise natural-language description of the spectrum generated by DeepSeek-V3.2, constrained to use only information present in the fact structure. The full generation prompt is shown in Figure~\ref{fig:alignment-prompt}.

\begin{figure}[H]
\centering
\begin{promptbox}
\footnotesize
You are an experienced astronomical spectroscopist. Your task is to write a natural and objective textual description based on the provided structured 1D spectral feature report. This description will be used to align a spectral encoder with a language model. Therefore, your goal is not to generate a measurement table or a lengthy physical interpretation, but to describe the observed spectral features in natural language while preserving key information useful for spectrum-language alignment, such as the object type, continuum morphology, main spectral lines, approximate observed positions of important lines, absorption or emission nature, relative strength, and reliable physical parameter context.

\medskip
\textbf{General style requirements:}
\begin{itemize}[leftmargin=*,nosep]
    \item Do not mention JSON, field names, structured reports, confidence labels, measurement tables, or phrases such as ``the report shows'', ``the data indicate'', ``the input information shows'', or ``the feature analysis shows''.
    \item The description should sound like a natural observational description written by an astronomical spectroscopist, rather than a sentence-by-sentence conversion of a measurement table. Never use ``identifies'', ``identified as'', ``classifies'', or similar classifier-like wording.
    \item For stellar spectra, prefer describing radial velocity or the blueshift/redshift of stellar lines, rather than treating redshift as the central quantity.
    \item The continuum should only be described as an observed spectral morphology. Describe its color, slope, or overall shape, such as blue, red, flat, or smoothly varying, only when the continuum information is reliable.
    \item For spectral features that are explicitly indicated as significant, strong, or visible, you may use modifiers such as clear, strong, prominent, or visible. You may naturally include their observed wavelengths, absorption or emission nature, and approximate equivalent widths. However, do not mechanically list detailed information for every spectral line.
    \item For weak or uncertain spectral features, use cautious wording such as weak, faint, shallow, subtle, or tentative.
    \item Mention the 4000\,\AA\ break only when it is genuinely useful for describing the object type and spectral appearance. For QSOs, do not emphasize the 4000\,\AA\ break unless it is clearly important.
    \item When physical parameters are reliable, you may naturally incorporate $T_{\mathrm{eff}}$, $\log g$, [Fe/H], redshift, or radial velocity, but do not mechanically list all available parameters.
    \item Avoid excessive physical interpretation or over-explaining the physical mechanisms.
    \item Ensure that the description is scientifically accurate and suitable for spectrum-language model alignment.
\end{itemize}

\end{promptbox}
\caption{Full prompt used to generate spectrum--caption pairs for alignment pretraining.}
\label{fig:alignment-prompt}
\end{figure}

\clearpage
\section{Additional Experimental Details}
\label{app:experimental_details}
\subsection{Training Hyperparameters}
\label{app:training-details}

Table~\ref{tab:training-hyperparams} summarizes the main optimization hyperparameters used for SpecMAE pretraining and AstroSpecLM spectral instruction tuning.

\begin{table}[H]
\centering
\small
\begin{tabular}{lcc}
\toprule
\textbf{Hyperparameter} & \textbf{SpecMAE} & \textbf{AstroSpecLM} \\
\midrule
Training data & \(\mathcal{D}_{\mathrm{enc}}\) & DESI-SpecInstruct \\
Epochs & 100 & 1 \\
Batch size & 360 & 32 \\
Optimizer & AdamW & AdamW \\
Learning rate & \(1\times10^{-4}\) & \(2\times10^{-5}\) \\
Weight decay & 0.001 & 0.01 \\
Scheduler & Cosine decay & Cosine decay \\
Warmup & 5 epochs & 100 steps \\
Precision & bf16 & bf16 \\
Training time & \(\sim\)24 hours & \(\sim\)3 hours \\
\bottomrule
\end{tabular}
\caption{Training hyperparameters for SpecMAE pretraining and AstroSpecLM spectral instruction tuning.}
\label{tab:training-hyperparams}
\end{table}

\subsection{Explanation Faithfulness Evaluation}
\label{app:faithfulness_prompt}

We use DeepSeek-V3.2 as an automatic judge to evaluate explanation faithfulness. The judge receives the generated explanation and the corresponding held-out spectral facts, but not the ground-truth final answer label. It assigns a score from 1 to 5, where 1 indicates an unsupported or hallucinated explanation and 5 indicates that the explanation is fully supported by the spectral facts. The full judging prompt is shown in Figure~\ref{fig:faithfulness-judge-prompt}.

\begin{figure}[H]
\centering
\begin{promptbox}
\footnotesize
You are an expert astrophysicist and a rigorous data validation specialist. Your task is to evaluate whether a spectrum-language model's generated natural-language explanation is faithful to the reference spectral facts and whether it contains any hallucinations.

\medskip
You will be provided with:
\begin{itemize}[leftmargin=*,nosep]
 \item \textbf{Reference Facts:} catalog-derived physical parameters and measured spectral features used as the evaluation reference.
 \item \textbf{Question:} the user's query or instruction.
 \item \textbf{Model Answer:} the generated explanation to be evaluated.
\end{itemize}

Please objectively evaluate the \textbf{Model Answer} strictly based on the \textbf{Reference Facts}. Do not use external knowledge that conflicts with the provided facts.

\medskip
\textbf{Evaluation Dimensions:}
\begin{itemize}[leftmargin=*,nosep]
    \item \textbf{Hallucination:} Determine whether the model fabricates features not present in the Reference Facts (e.g., inventing non-existent emission or absorption lines) or generates statements that contradict the facts (e.g., stating an incorrect redshift or object class).
    \item \textbf{Faithfulness Score (1--5 Scale):} Assess how well the model's answer aligns with the given facts:
    \begin{itemize}[leftmargin=*,nosep]
        \item \textbf{1 (Severe):} Completely contradictory or filled with severe misleading hallucinations.
        \item \textbf{2 (Poor):} Partially correct, but contains major factual errors or critical fabrications.
        \item \textbf{3 (Fair):} Mostly correct, but includes minor numerical deviations or trivial fabricated details that do not alter the main physical interpretation.
        \item \textbf{4 (Good):} Highly accurate and faithful. No obvious errors, though it may omit some minor facts.
        \item \textbf{5 (Excellent):} Perfectly accurate. Flawlessly covers the key facts without any hallucinations or contradictions.
    \end{itemize}
\end{itemize}

\medskip
\textbf{Output Format:} \\
You must output your evaluation strictly in JSON format with the following keys:

\begin{verbatim}
{
  "reasoning": "...",
  "has_hallucination": true or false,
  "hallucination_type": "None" or "Contradiction" or "Fabrication",
  "faithfulness_score": 1--5
}
\end{verbatim}

\end{promptbox}
\caption{Full prompt used for the LLM-as-a-judge faithfulness evaluation.}
\label{fig:faithfulness-judge-prompt}
\end{figure}
\end{document}